\documentclass[10pt,letterpaper]{article}

\usepackage{times}
\usepackage{color}
\usepackage{amsmath}
\usepackage{amssymb}
\usepackage{subfig}
\usepackage{wrapfig}
\usepackage{graphicx}
\usepackage{float}
\usepackage{makecell}
\usepackage[format=plain,justification=centering]{caption}
\usepackage[backend=biber, style=ieee]{biblatex}
\usepackage[breaklinks=true,colorlinks,bookmarks=false]{hyperref}
\newcommand{\ignorethis}[1]{}

\title{Classifying High-Energy Celestial Objects with Machine Learning Methods}

\author{
    Nolan Faught \and
    Tyrian Hobbs \and 
    Alexis Mathis \and 
    Daniel Yu \and
Department of Mathematics, Northeastern University
}

\begin{document}

\maketitle

\begin{abstract}
Machine learning is a field that has been growing in importance since the
early $2010$s due to the increasing accuracy of classification models
and hardware advances that have enabled faster training on large datasets.
In the field of astronomy, tree-based models and simple neural networks
have recently garnered attention as a means of classifying celestial
objects based on photometric data. We apply common tree-based models to
assess performance of these models for discriminating objects with
similar photometric signals, pulsars and black holes.

We also train a RNN on a downsampled and normalized version of the
raw signal data to examine its potential as a model capable of object
discrimination and classification in real-time.
\end{abstract}

\section{Introduction}\label{sec:intro}
Modern astronomy has generated an extensive taxonomy of celestial objects based on their physical characteristics and predicted future state. As theories
of the development, expansion, history, and predicted future state of the universe rely on identifying and observing celestial bodies, it is essential to have quick and accurate classification of newly observed objects. Historically, classification was performed manually, but the rapid expansion of modern
catalogues of celestial objects -- such as the Sloan Digital Sky Survey, which grows at a rate of thousands of entries daily \cite{york2000sloan} -- makes this manual classification impractical.

Supervised and semi-supervised machine learning represent the most promising candidates for the desired computational classification. Until recently, the data, hardware, and software required for large-scale training and deployment of these methods were unavailable to the general research community. However, improvements
to parallel processing hardware have driven increased success and adoption,
resulting in the invention of models capable of equaling or surpassing
human-level intelligence in tasks formerly considered intractable
to computers. Such improvements have been recognized in facial recognition \cite{parkhi2015deep} and
combinatorial game theory \cite{hsu1995deep}, but despite their meteoric rise in
popularity, there is a significant gap in astronomical literature on applying
machine learning models to the problem of celestial object classification.

In an effort to improve this state, we explore a number of machine learning based models for a
simplified celestial object classification problem to assess the performance
and potential of these models in the field of astronomy.

\section{Literature Review}\label{sec:survey}
Previous work has applied convolutional neural networks \cite{harshvardhan2024}
and tree-based models to classification problems in broader categories
\cite{zeraatgari2023} using photometric data. Models that rely on features extracted from photometric data, such as gravitional tensors, have also
gained attention \cite{10928692}, \cite{barapatre2025}. However, there seems to be a gap in the literature in applying tree-based models to discriminate more photometrically
similar object classes, such as black holes and pulsars.

This work attempts to address this gap by examining both previously considered computational discrimination models in a more specific task environment to address their performance, and additionally explores a recurrent neural network model trained directly on downsampled photometric digital signals with the intention of creating a model capable of real-time
inference.

\section{Background}\label{sec:background}
Celestial objects are classified by the characteristics of
their emitted and reflected light. By examining the spectra and patterns of
the light curve -- the time-varying signal of frequency and
intensity from a single object -- it is possible to determine the mass, momentum, rotation, gravitational field, and composition of an object with high accuracy and precision. Modern high-powered telescopes record digital signals in both high- and low-frequency bands of the electromagnetic spectrum, providing viable data to characterize these objects.

Two particular celestial objects are difficult to discriminate due to their similar photometric signals: black holes and pulsars. A black hole is
a celestial object so dense that nothing -- even light -- can escape from its gravitational pull. What we view is the accretion disk, a layer of rapidly-spinning material that is heated by friction and shines brightly. A pulsar is a rotating neutron star that emits jets of matter and electromagnetic radiation from its magnetic poles. Much like a lighthouse, the rotation of the neutron stars
periodically sweeps these cones of light in the direction of the Earth, causing
us to see a pulsating star, or a pulsar. These pulses occur at very regular intervals, typically ranging from milliseconds to seconds. 

Theoretically, the electromagnetic signal received from a black hole is constant, with occasional spikes due to jets of escaped material. Pulsars, on the other hand, should exhibit a sinusoidal light curve with regular spikes corresponding to the times the pulsar's magnetic poles are facing the earth. In practice, owing to cosmic noise and other visual factors, these two objects can be difficult to distinguish without careful consideration.

\begin{figure}[htpb]
    \centering
    \begin{minipage}{0.3\textwidth}
        \centering
        \includegraphics[width=\textwidth]{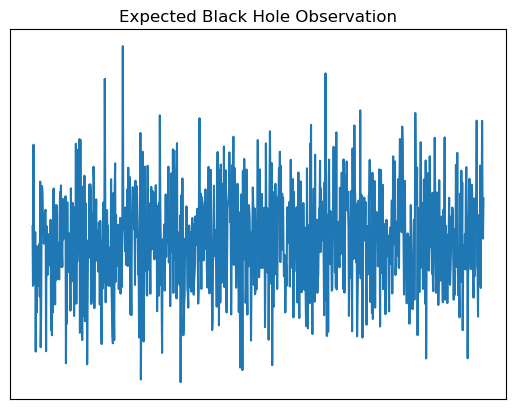}
    \end{minipage}
    \hspace{3cm}
    \begin{minipage}{0.3\textwidth}
        \centering
        \includegraphics[width=\textwidth]{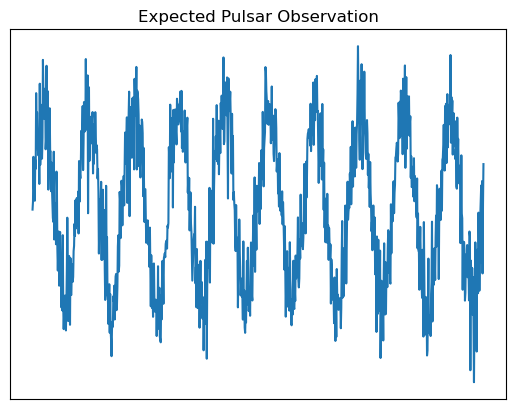}
    \end{minipage}
    \caption{Ideal Light Curves}
    \label{fig:idealcurves}
\end{figure}

Our models are trained on the photometric signals of black holes and pulsars
in the hard X-ray band, a high-frequency band of the EM spectrum. These data
are all provided from the same instrument, the Nuclear Spectroscopic Telescope
Array (NuSTAR), a NASA telescope which has been performing observational
research on a variety of objects in the X-ray wavelength since 2012.
\cite{nustar_website} It has greater sensitivity compared to previous instruments
operating in similar energy bands, making it particularly useful for study of
celestial objects.

\section{Methodology}\label{sec:methods}

\begin{wrapfigure}[13]{r}{0.3\textwidth}
    \includegraphics[width=0.3\textwidth]{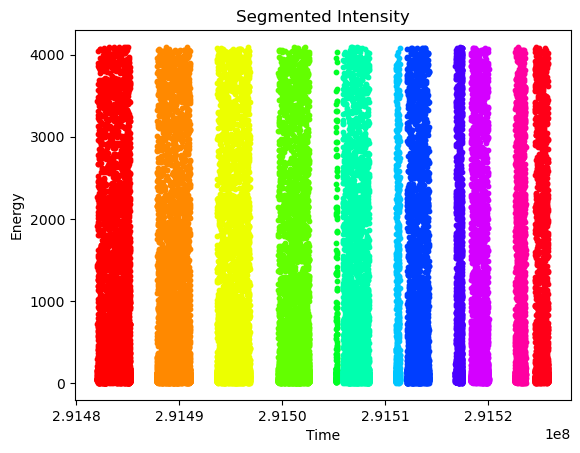}
    \caption{Gaps in event arrival times (colored by cluster)}
\end{wrapfigure}

Target sources for this study included supermassive black holes and millisecond
pulsars selected based on their observational characteristics and scientific
interest. Data were acquired via the High Energy Astrophysics Science Archive
Research Center (HEASARC) browsing system by querying the NuSTAR mission master
database for all publicly available observations of the selected targets; the full list of objects viewed and their number of suitable observations is listed in the appendix. For
each observation meeting selection criteria, cleaned event files were retrieved from the archive.

The event format (\texttt{.evt}) is a file format common to astronomical data used to store a telescope's recorded output. Necessary for this analysis was the pixel location, UTC time, and pulse-invariant (PI) channel of incoming events (photons). NuSTAR provides event files that have been processed to remove sources of measurement error such as cross-correlation between pixels and shifts to the spectra caused by exposure variations, as documented in the NuSTAR software guide \cite{perri2013nustar}.

NuSTAR's cleaning process also includes removal of ``dirty'' data; this may occur due to instrument dead time, telescope movement, the satellite passing through high-radiation sections of space (such as the South Atlantic Anomaly), or any other event that may damage data quality. Thus, our observations contained large gaps between portions of usable data, often around three thousand seconds in length.

Otherwise, events occurred at $0.25$s to $0.65$s intervals, with a median wait time of $0.5$s. A histogram analysis revealed that arrival times appeared to follow a Poisson distribution with an approximate rate of $0.5$s. 

\subsection{Preprocessing and Normalization}

\begin{wrapfigure}[19]{l}{0.3\textwidth}
    \centering
    \includegraphics[width=0.3\textwidth]{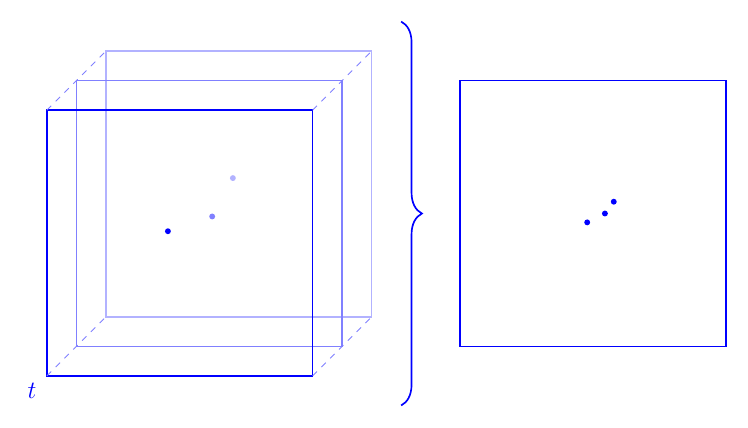}
    \includegraphics[width=0.3\textwidth]{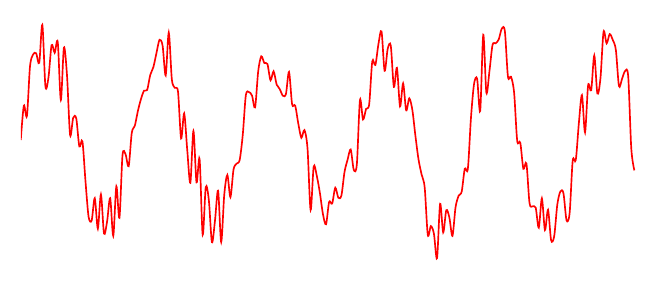}
    \caption*{\emph{Top:} Integration of sparse signal matrices over time \\
              \emph{Bottom:} Real-valued signal at a constant period}
\end{wrapfigure} 

The digital signals were reconstructed as dense
$k \times n \times n$ real-valued tensors, where $k$ represented event time and $n\times n$ represented the spatial aperture (camera pixels), zero-padded to a uniform size. An initial reconstruction approach consisted of integrating photon arrivals at constant periods with $T \in \{0.5, 1, 5, 15, 30\}$ second intervals, but the resulting
signals proved too sparse to perform analysis. Spatial downsampling, filtering, and smoothing methods were considered as potential alternatives, but ultimately rejected as outside the scope of this work.

Since there is no expected difference in shape between the two celestial objects, the spatial characteristics of photon arrivals were ultimately discarded. Instead, the energy of the events were integrated over time to create a real-valued signal with a constant period of one second. This period was selected after experimentation as the shortest interval capable of producing signals with sufficient density for analysis. The resulting signals were first segmented at the aforementioned data gaps, then processed with a sliding-window approach with thirty minute (1800s) windows and a thirty second stride. 

For feature-based models, analysis of the $k$ largest Fourier coefficients was considered, but again proved outside the scope of this research. Instead, ten statistical features were extracted from each processed intensity signal. These consisted of sample mean, the five quartile values, standard deviation, variation, skewness, and kurtosis.

\subsection{Data Exploration}

\begin{wrapfigure}[11]{r}{0.14\textwidth}
    \vspace{-10pt}
    \includegraphics[width=0.14\textwidth]{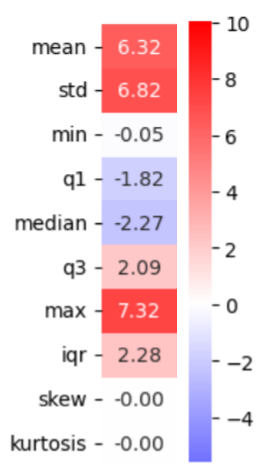}
    \caption{Average Statistic Differences}
    \label{fig:placeholder}
    \vspace{-20pt}
\end{wrapfigure} 

The finalized intensity dataset consisted of 1800 features, one for each second of the thirty-minute (1800s) observation interval, and the target label (pulsar or black hole). There was an approximate 3:1 class imbalance, with black hole observations significantly outnumbering pulsars.

\begin{figure}[b!]
    \centering
    \begin{minipage}{0.25\textwidth}
        \centering
        \includegraphics[width=\textwidth]{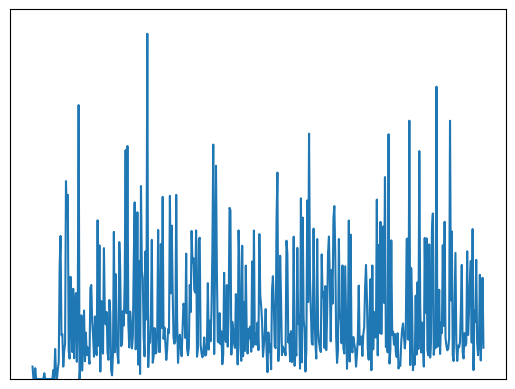}
        \caption{Sample Black Hole}
    \end{minipage}
    \hspace{3cm}
    \begin{minipage}{0.25\textwidth}
        \centering
        \includegraphics[width=\textwidth]{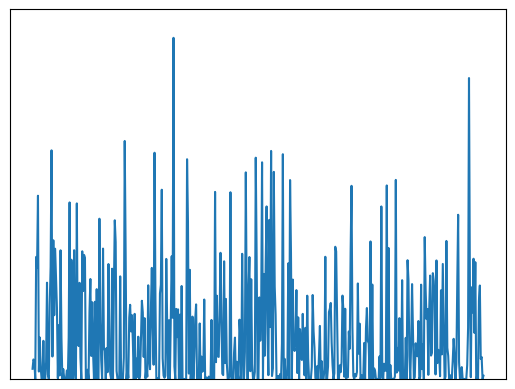}
        \caption{Sample Pulsar}
    \end{minipage}
\end{figure}

The figure below shows sample time series of a randomly selected black hole and pulsar. Compared to the ideal theoretical distributions, the observed sample distributions are notably difficult to distinguish; there are no significant differences in amplitude distribution, cycle length, or baseline drift that could potentially allow for class separation through visual inspection or linear separation. This similarity demonstrates the utility of machine intervention. 

Examining the summary statistics described above, second-order summaries were computed across their distribution, and compared between the two classes. The difference between the averages for each feature is visible on the right, alongside the logarithmic scale.

Pulsars demonstrated on average far higher means, standard deviations, and max values, and had lower first quartiles and medians. Skewness and kurtosis showed no major differences; this is unsurprising, as both were on a very small scale, and the intensity curves were generally cyclic.

These findings indicate that pulsars produce infrequent but strong high-amplitude pulses, while their median behavior remains more muted.  

\subsection{Models}

Following from the general tendencies present in previous literature, three models were trained based on the statistical features: a logistic regression model, a random forest model, and XGBoost.
These models were trained with cloud computing resources provided by the collaborative scientific research platform SciServer.

Finally, a recurrent neural network was trained directly on the downsampled signal data. While convolutional and deep neural networks were considered, the
time-dependent nature of the data demonstrated a clear suitablity for recurrent neural networks. Due to computational limitations of SciServer, this
model was trained on physical hardware, as discussed further below.

Since the dataset contained a class imbalance favoring black holes over pulsars, training and validation data was stratified to help account for this disparity.

\subsubsection{Logistic Regression}

Logistic regression models the log-odds of the class probability as a linear combination of the input features. Because the model is inherently linear, it cannot capture nonlinear patterns or feature interactions unless nonlinear or interaction features are explicitly constructed and provided as inputs.

The logistic regression was trained for $1000$ iterations.

\subsubsection{Random Forest}

The second model trained was a random forest, created using the summary statistics dataset and the scikit-learn RandomForestClassifier module. 

RandomForestClassifier trains decision trees based on random subsets of the available features, then aggregates their outputs to create a final decision. Our training created one hundred such decision trees, utilizing three randomly chosen features to make each splitting decision. Decisions were made based on improving the Gini purity of bootstrapped samples the same size as the initial dataset. We did not set a max tree depth, instead allowing the decision tree to continue splitting until the nodes were completely pure.

\subsubsection{XGBoost}
\begin{wrapfigure}[9]{r}{0.3\textwidth}
    \vspace{-10pt}
    \includegraphics[width=0.3\textwidth]{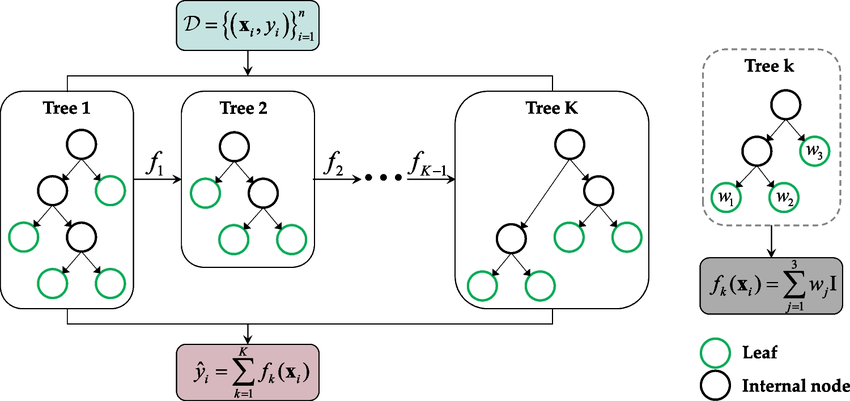}
    \caption{XGBoost Architecture}
    \label{fig:placeholder}
\end{wrapfigure} 

The third model trained was an XGBoost (eXtreme Gradient Boosting) classifier, which builds an ensemble of decision trees sequentially. Each new tree is trained on the gradient (and second-order gradient) of the loss with respect to the current model’s predictions, allowing it to correct the residual errors made by the previous trees.

For our model, we decided to use a standard learning rate of $0.05$ and a sequence of $400$ trees with each tree at a maximum depth of $10$ branches. In order to minimize overfitting, we randomly sample only $80$ percent of the data and 80 percent of the features for each tree created in the ensemble. Similarly, we also re-weight the imbalanced classes.

This combination of parameters were chosen to prevent overfitting while maximizing accuracy.

\subsubsection{RNN}

Our most complicated model was a bidirectional Long Short-Term Memory (LSTM)
recurrent neural network (RNN) trained on photometric signals normalized with
z-score scaling. 

\begin{wrapfigure}[11]{l}{0.3\textwidth}
    \vspace{-10pt}
    \includegraphics[width=0.3\textwidth]{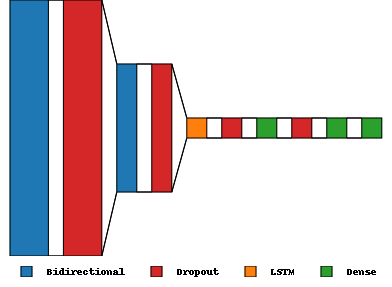}
    \caption{RNN Architecture}
\end{wrapfigure} 

The structure of the RNN contains ten layers: 2 bidirectional LSTM,
1 standard LSTM, 3 dense, and 4 dropouts. Multiple LSTM layers were used to
increase the opportunity that the RNN would extract the important features of the
signal as simpler signals. Bidirectional LSTM layers were included due to their
effectiveness at distinguishing features of periodic signals. Dropout layers
turned off $30\%$ neurons three times and $40\%$ of neurons once during the third iteration of training. Two of the dense layers used a ReLU activation function, and
the final layer a sigmoid activation.

The model was trained using the Adam optimizer with binary cross-entropy loss.
Training, test, and validation performance were assessed with model accuracy.
Due to the volume of data and temporal requirements for training
the model, this model was trained on a physical desktop with a $12$-core CPU
and $64$GiB RAM. Training took about $11$ continuous hours of operation at
$1.2$ CPU core usage, requiring $24$GiB of memory.

The RNN model was trained twice on subsets of the data. The first training
period was performed on 5 objects and took about $2$ hours to complete.
The results of this training period were trained on $60\%$ of the
remaining data.

\section{Results}

\subsection{Logistic Regression}

\begin{wrapfigure}[12]{r}{0.3\textwidth}
    \vspace{-10pt}
    \includegraphics[width=0.3\textwidth]{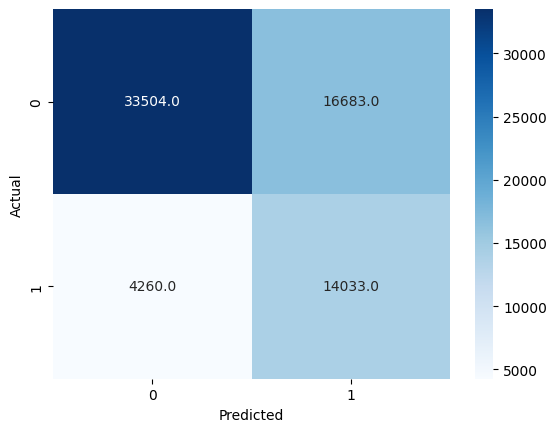}
    \caption{Logistic Regression Confusion Matrix}
    % \vspace{-10pt}
\end{wrapfigure}

Logistic regression was adopted as the baseline model and attained a training accuracy of $0.711$ and a test accuracy of $0.71$, indicating the model has generalized well from training to test data. However, the confusion matrix shows that the model is misclassifying a significant number of black holes as pulsars. This tradeoff for increased sensitivity to the minority class, pulsars $(1)$, at the cost of reduced accuracy for the majority class, black holes $(0)$, is explained by the re-weighting of the model to correct for the class imbalance.

The model’s inability to simultaneously improve accuracy on both classes suggests that the underlying decision boundary is not well-approximated by a linear separator. In such settings, logistic regression exhibits high model bias and cannot capture the complex, nonlinear structure present in the data. Thus models such as Random Forest or XGBoost should result in improved performance.

\subsection{Random Forest}

\begin{wrapfigure}[11]{l}{0.3\textwidth}
    \vspace{-10pt}
    \includegraphics[width=0.3\textwidth]{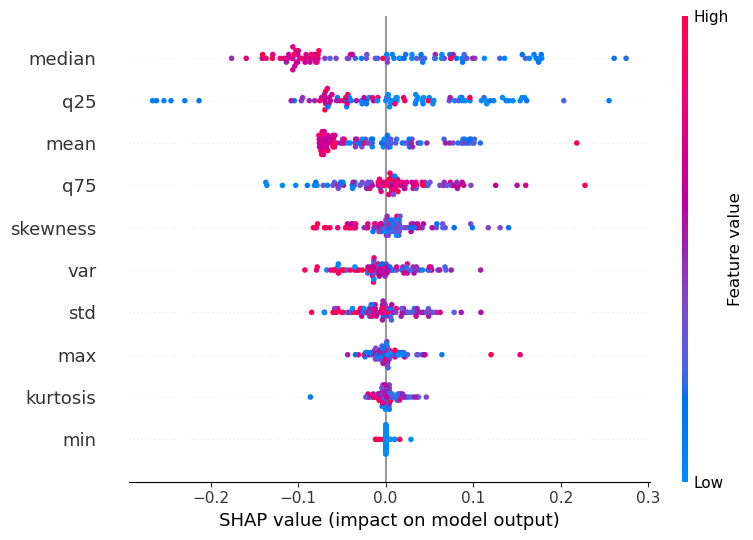}
    \caption{Random Forest SHAP values}
\end{wrapfigure}

The random forest model achieved an initial training accuracy of $1$, which is not uncommon for random forest models in general. The random forest model achieved a total test accuracy of $0.931$. It had precision of 0.94 and recall of 0.97 with black holes, and precision of 0.90 and recall of 0.84 with pulsars. It assigned highest importance to the median with a feature importance value of 0.1873, followed by the 25th quartile and the mean with feature importances $0.1523$ and $0.13$ respectively. 

Unfortunately, this also came with an average tree depth of $42.7$ splits (minimum $38$, maximum $53$). These large trees increase computational time, reduce interpretability, and while the random nature of the training helps combat overfitting it is still a potential problem. Nevertheless, this is a promising result that suggests more exploration. 
 
\subsection{XGBoost}

XGBoost achieved $0.938$ accuracy on the training set and $0.92$ accuracy on the
test set, similar in accuracy to the random forest model. Trees were shallower
compared to the random forest, which we conjecture has resulted in less overfitting.

Examining the confusion matrix, the model had a slight tendency to misclassify black holes towards pulsars. However, this is not significant and accuracy in both pulsar and black hole classification improved compared to the logistic regression model, indicating the model picked up on nonlinear interactions that logistic regression was unable to.

The SHAP values coincide with the values for the Random Forest model. The top two features are identical and three out of five top features are the same. The features picked out as important all describe the central tendency and tail behavior of the distribution, which is consistent with the analysis during data exploration. Specifically, the models agree that pulsars tend to have higher medians, shaper peaks, and greater variance. However, while the models pick up similar non linear patterns, XGBoost seems to have more outliers in terms of model impact than the random forest.

\begin{figure}[htbp]
    \centering
    
    \begin{minipage}{0.30\linewidth}
        \centering
        \includegraphics[width=\linewidth]{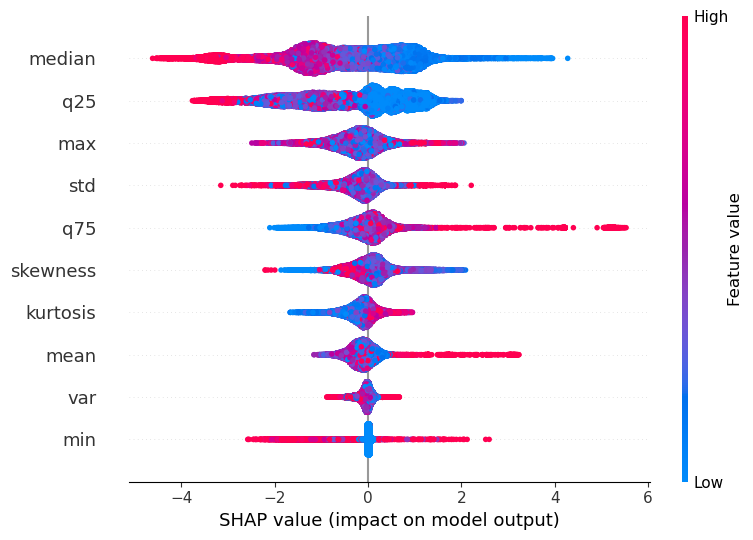}
        \caption{XGBoost SHAP Values}
    \end{minipage}
    \hspace{3cm}
    \begin{minipage}{0.30\linewidth}
        \centering
        \includegraphics[width=\linewidth]{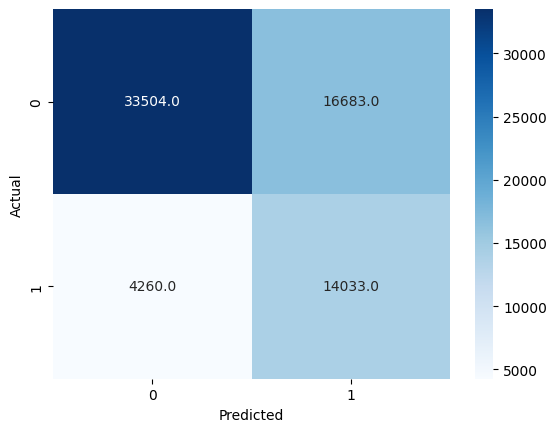}
        \caption{XGBoost Confusion Matrix}
    \end{minipage}
\end{figure}

\subsection{Recurrent Neural Network}

%\begin{figure}[H!]
  %  \vspace{-10pt}
  %  \includegraphics[width=0.3\textwidth]{Figures/smalldset_confussion.png}
 %   \caption{RNN confusion matrix for small dataset with ROC curve}
 %   \label{fig:placeholder}
%\end{figure}

\begin{wrapfigure}[13]{l}{0.3\textwidth}
    \vspace{-10pt}
    \includegraphics[width=0.3\textwidth]{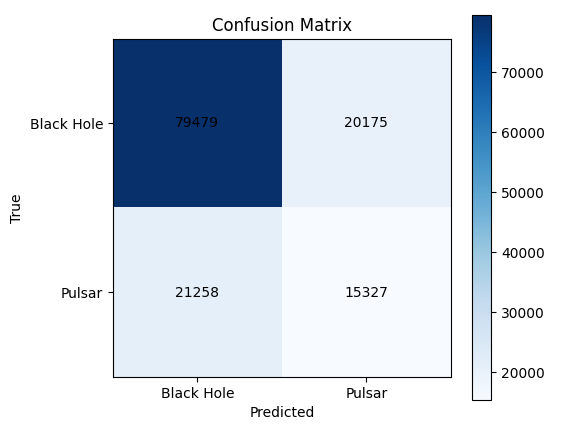}
    \caption{RNN confusion matrix for large dataset}
    \label{fig:placeholder}
\end{wrapfigure}

\begin{figure}[b!]%
    %\centering
    %\subfloat[\centering Loss, accuracy, and AUC of the training and validation for the small dataset]{{\includegraphics[width=5cm]{Figures/smalldset_three.png} }}%
    %\qquad
    \centering
    \includegraphics[width=15cm]{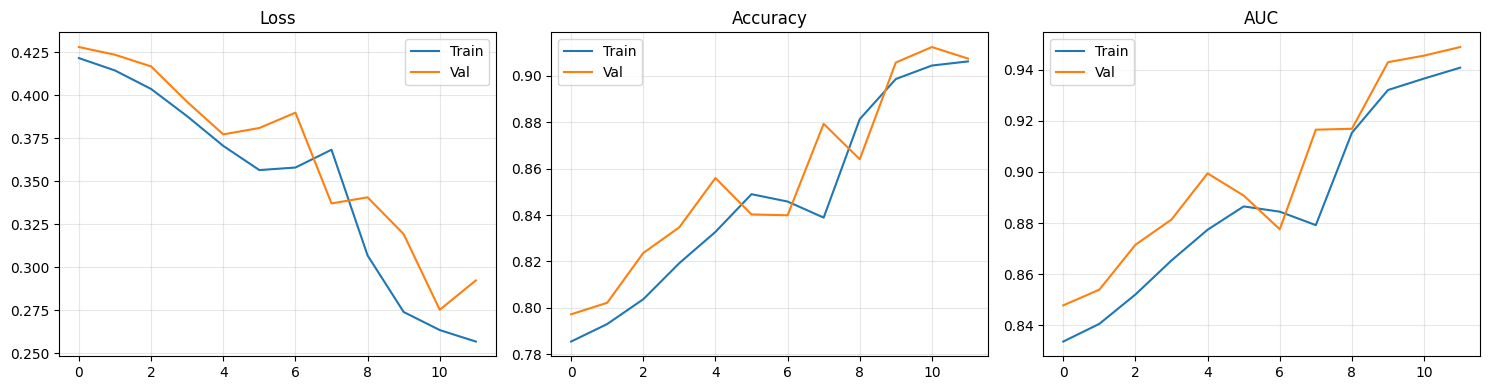}%
    \caption{Loss, accuracy, and AUC of the training and validation}%
    \label{fig:example}%
\end{figure}

The recurrent neural network demonstrated high training accuracy at $0.91$
but low test accuracy at $0.69$. This is likely an artifact of overfitting, but
due to time constraints we were unable to retrain the model or attempt to
refine the training process. Adjustments to the structure or data pre-processing
that may improve results are discussed in section 6.1.

\clearpage
\subsection{Model Cross-Comparison}

\begin{table}[htpb]
    \centering
    \begin{tabular}{|c|c|c|c|c|}
        \hline
        Model & Training Accuracy & Test Accuracy  & Training Time\\
        \hline
        Logistic Regression & 0.711 & 0.71 & 4m 24.62s\\
        Random Forest & 1.00 & 0.931 &  1m 6.12s \\
        XGBoost & 0.938 & 0.92 & 1m 1.18s \\
        RNN & $0.91$ & $0.69$ & 10hr 49m \\
        \hline
    \end{tabular}
\end{table}

\section{Discussion}

Tree-based models had comparable accuracy and precision to other models in
the literature, which shows that tree-based models are apt for more specific
celestial object discrimination. The simplicity of our data pre-processing and
lack of signal processing are a weakness of our approach, so more accurate
and mathematically rigorous models should demonstrate improved performance.

The accuracy of the recurrent neural network showed steady improvement prior
to stopping the training period. Tuning the downsampling or combining spatial
aspects of the signal to produce a combined recurrent-convolutional neural
network may be viable approaches for real-time signal processing, but the
current results are inconclusive.

Our results show that a purely ML approach to classifying pulsars and black holes is indeed viable and can classify pulsars and black holes to a high degree of accuracy. 

\subsection{Further Work}
The use of statistical methods for digital signals may be a significant source
of error due to cross-correlation of features in time. The mean and median, in
particular, are unreliable for discrimination problems due to correlating directly
to distance from the Earth.

Examining the tuning, training, and structure of the recurrent neural network proved
too computationally intensive for in-depth analysis. Improvements that we considered
include mostly the pre-processing and cleaning of signals, such as spatial
filtering, clutter removal, spatial/temporal smoothing, and larger intervals
between signal sampling. 

\section{Acknowledgments}
This research has made use of data, software and/or web tools obtained from the
High Energy Astrophysics Science Archive Research Center (HEASARC), a service of
the Astrophysics Science Division at NASA/GSFC and of the Smithsonian
Astrophysical Observatory's High Energy Astrophysics Division.
 
This research made use of the SciServer science platform (www.sciserver.org).
SciServer is a collaborative research environment for large-scale data-driven
science. It is being developed at, and administered by, the Institute for Data
Intensive Engineering and Science at Johns Hopkins University. SciServer is
funded by the National Science Foundation through the Data Infrastructure
Building Blocks (DIBBs) program and others, as well as by the Alfred P. Sloan
Foundation and the Gordon and Betty Moore Foundation.

This research has used programming assistance provided by Anthropic's Claude Sonnet 4.5 model.
All outputs were confirmed by a researcher and no AI assistance was used for
interpretation, review, or final reporting.

This research was largely possible due to the open-source programming library Scikit-learn. We would like to thank the creators and maintainers of that module for support of the academic community.

\nocite{*}
\printbibliography

@article{white2003heasarc,
  title={HEASARC Software Archive},
  author={White, Nicholas and Murray, Stephen S},
  year={2003}
}

@article{perri2013nustar,
  title={The NuSTAR data analysis software guide},
  author={Perri, M and Puccetti, S and Spagnuolo, N and Ficcadenti, R and Davis, A and Forster, K and Grefenstette, B and Harrison, F and Madsen, K},
  journal={ASI Space Science Data Center and California Institute of Technology (Julio 2021)},
  year={2013}
}

@INPROCEEDINGS{10928692,
  author={Siddiquee, Md Fairuz and Hasan, Md Mehedi},
  booktitle={2024 IEEE 3rd International Conference on Robotics, Automation, Artificial-Intelligence and Internet-of-Things (RAAICON)}, 
  title={Spectral and Morphological Classification of Celestial Objects using Physics Informed Machine Learning}, 
  year={2024},
  volume={},
  number={},
  pages={7-12},
  doi={10.1109/RAAICON64172.2024.10928692}
}

@article{ahmed2023,
author = {Hassina, Ahmed Taha},
journal = {viXra},
year = {2023},
month = {08},
pages = {},
title = {Using machine learning to classify and localize stellar objects}
}

@article{barapatre2025,
author = {Omprakash Barapatre, Shubham Sanskar Routray, Shruti Patel},
journal = {IJRTI},
year = {2025},
title = {A Physics-Informed Hybrid Machine Learning Pipeline for Celestial Object Classification}
}

@misc{zeraatgari2023,
      title={Machine learning-based photometric classification of galaxies, quasars, emission-line galaxies, and stars}, 
      author={Fatemeh Zahra Zeraatgari and Fatemeh Hafezianzade and Yanxia Zhang and Liquan Mei and Ashraf Ayubinia and Amin Mosallanezhad and Jingyi Zhang},
      year={2023},
      eprint={2311.02951},
      archivePrefix={arXiv},
      primaryClass={astro-ph.GA},
      url={https://arxiv.org/abs/2311.02951}, 
}

@inproceedings{parkhi2015deep,
  title={Deep face recognition},
  author={Parkhi, Omkar and Vedaldi, Andrea and Zisserman, Andrew},
  booktitle={BMVC 2015-Proceedings of the British Machine Vision Conference 2015},
  year={2015},
  organization={British Machine Vision Association}
}

@inproceedings{hsu1995deep,
  title={Deep Blue system overview},
  author={Hsu, Feng-hsiung and Campbell, Murray S and Hoane Jr, A Joseph},
  booktitle={Proceedings of the 9th international conference on Supercomputing},
  pages={240--244},
  year={1995}
}

@article{york2000sloan,
  title={The sloan digital sky survey: Technical summary},
  author={York, Donald G and Adelman, Jennifer and Anderson Jr, John E and Anderson, Scott F and Annis, James and Bahcall, Neta A and Bakken, JA and Barkhouser, Robert and Bastian, Steven and Berman, Eileen and others},
  journal={The Astronomical Journal},
  volume={120},
  number={3},
  pages={1579},
  year={2000},
  publisher={IOP Publishing}
}

@InProceedings{harshvardhan2024,
author="Gaikwad, Harshvardhan
and Mhala, Nikhil
and Umare, Atharva
and Milmile, Aniket
and Lanjewar, Aditya",
editor="Nanda, Umakanta
and Tripathy, Asis Kumar
and Sahoo, Jyoti Prakash
and Sarkar, Mahasweta
and Li, Kuan-Ching",
title="Classification of Star and Galaxy Objects Utilizing Machine Learning Techniques and Deep Neural Networks",
booktitle="Advances in Distributed Computing and Machine Learning",
year="2024",
publisher="Springer Nature Singapore",
address="Singapore",
pages="13--23",
isbn="978-981-97-3523-5"
}

@article{scikit-learn,
  title={Scikit-learn: Machine Learning in {P}ython},
  author={Pedregosa, F. and Varoquaux, G. and Gramfort, A. and Michel, V.
          and Thirion, B. and Grisel, O. and Blondel, M. and Prettenhofer, P.
          and Weiss, R. and Dubourg, V. and Vanderplas, J. and Passos, A. and
          Cournapeau, D. and Brucher, M. and Perrot, M. and Duchesnay, E.},
  journal={Journal of Machine Learning Research},
  volume={12},
  pages={2825--2830},
  year={2011}
}

@misc{nustar_website,
    author = {{NuSTAR Science Operations Center}},
    title = {{NuSTAR -- Nuclear Spectroscopic Telescope Array}},
    howpublished = {\url{https://nustar.caltech.edu/}},
    note = {Accessed: 2025-12-09},
    year = {2025}
}

\clearpage
\section{Appendix}

{\begin{table}[H]
    \centering
    \begin{tabular}{|l|l|c|}
        \hline
        Object Name & Object Type & \makecell{Number \\ of \\ Observations} \\
        \hline
        3C 264               & Black Hole & 2 \\
        APM 08279            & Black Hole & 4 \\
        Centaurus A*         & Black Hole & 12 \\
        M87*                 & Black Hole & 38 \\
        MCG-6-30-15          & Black Hole & 12 \\
        NGC 1275             & Black Hole & 4 \\
        NGC 4151             & Black Hole & 26 \\
        NGC 4388             & Black Hole & 2 \\
        NGC 4579             & Black Hole & 2 \\
        Sag A*               & Black Hole & 152 \\
        IC10                 & Pulsar     & 6 \\
        IGR J00291           & Pulsar     & 2 \\
        M82 X2               & Pulsar     & 16 \\
        NGC 2808, MAXIJ0911  & Pulsar     & 6 \\
        NGC 5907             & Pulsar     & 30 \\
        NGC 6397             & Pulsar     & 2 \\
        NGC 6440, IGRJ 17488 & Pulsar     & 2 \\
        NGC 6441, 4U 1746    & Pulsar     & 2 \\
        NGC 6626, PSR B1821  & Pulsar     & 10 \\
        PSR B0833            & Pulsar     & 2 \\
        Terzan 5             & Pulsar     & 2 \\
        \hline
    \end{tabular}
    \caption{Celestial Objects acquired from HEASARC}
\end{table}}

\end{document}